\documentclass[12pt]{article} 
\usepackage[cp1251]{inputenc}
\usepackage[english]{babel}

\usepackage{amsmath}
\usepackage{amsthm}
\usepackage{amssymb}
\usepackage[left=1cm,right=1cm,top=2cm,bottom=2cm,bindingoffset=0cm]{geometry}
\usepackage{dsfont}
\usepackage{bm}
\usepackage{leftidx}
\usepackage[makeroom]{cancel}
\usepackage{indentfirst} 
\usepackage{hyperref}
\usepackage{tikz}
\usepackage{tikz-cd}
\usepackage[titletoc]{appendix}
\usetikzlibrary{matrix,arrows,decorations.pathmorphing}

\newcommand\abs[1]{\left|#1\right|}

\newcommand \op[1] {\hat{\bm{#1}}}

\DeclareMathOperator{\Span}{span}
\newcommand{\bma} {\begin{pmatrix}}
\newcommand{\ema} {\end{pmatrix}}
\newcommand{\Jnp}{\hat{J}_n^+}
\newcommand{\Jnm}{\hat{J}_n^-}
\newcommand{\Jnn}{\hat{J}_n^0}

\appendixtitleon

\makeatother

\author{
  {\Large{Michael Kreshchuk}}\\
  {\small{School of Physics and Astronomy, University of Minnesota}}\\
  {\small Minneapolis MN 55455 USA}\\
  {\small{e-mail: ~kreshchu @ physics.umn.edu~}}
}

\title{
{\Large{\textbf{
A quasi-exactly solvable model: two charges in a magnetic field, subject to a non-Coulomb mutual interaction\\}}}
}

\date{}

\begin{document}
\maketitle
\begin{abstract}
We extend the class of QM problems which permit for quasi-exact solutions. Specifically, we consider planar motion of two interacting charges in a constant uniform magnetic field. While Turbiner and Escobar-Ruiz (2013) addressed the case of the Coulomb interaction between the particles, we explore three other potentials.
We do this by reducing the appropriate Hamiltonians to the second-order polynomials in the generators of the representation of $SL(2,C)$ group in the differential form.
This allows us to perform partial diagonalisation of the Hamiltonian, and to reduce the search for the first few energies and the corresponding wave functions to an algebraic procedure.
\end{abstract}
\section{Introduction}

In quantum mechanics, quasi-exactly solvable (QES) problems are those which permit analytic derivation of a finite part of the (generally, infinite) spectrum. They allow for partial reducing of the Hamiltonian to a block-diagonal form. In these cases, one can find nonperturbative solutions for Hamiltonians of some special types. These special types are obtained from Hamiltonians of broader (generally, unsolvable) classes by constraining the entering parameters.

Exploration of these systems was pioneered by Turbiner \cite{turb1988}. Later, this kind of systems was independently rediscovered by Kamran and Olver \cite{Olver}. A different  approach (often termed {\it{analytic}}) was developed by Ushveridze \cite{UshP}. Various approaches to QES problems and some advances along this line of research are presented in~\cite{UshB}.
Among the recent advances in this field, notable are the discovery of `semiclassically quasi-exactly solvable potentials'~\cite{Semi}, as well as the results on studying elliptic~$A_n$ and~$BC_n$ models~\cite{A, BC}.
In the current paper, we employ a so-called algebraic approach, a detailed description whereof can be found in~\cite{Shifman}.

All exactly and quasi-exactly solvable one-dimensional Hamiltonians known so far can be reduced to second-order polynomials in the generators of a representation of the $SL(2,C)$ group:

\begin{equation}
\nonumber
\hat{h} = a_{\alpha \beta} \hat{J}^\alpha_n \hat{J}^\beta_n + b_\alpha \hat{J}^\alpha_n \quad,
\end{equation}
where~$\,\hat{J}^\alpha_n\,$ are implemented as differential operators. Such Hamiltonians' eigenvectors and eigenvalues are found through an algebraic procedure. Our goal is to broaden the class of problems which can be cast into the said form and which can, therefore, be treated by algebraic methods.
A list of QES Hamiltonians of the form
\begin{equation}
   \hat{H} = - \frac{d^2}{dx^2} + V(x)
   \label{vavila}
\end{equation}
was provided in~\cite{turb1988}.
When one considers the radial part of the equation for the relative motion in the problem of two particles in two or three dimensions, other QES Hamiltonians emerge. In distinction from the type~(\ref{vavila}), they contain the first derivative.
In~\cite{taut94}, several particular analytic solutions were described for a system of two electrons (interacting with Coulomb potentials) in a homogeneous magnetic field and an external oscillator potential. In~\cite{taut99}, analytic solutions were found for two particles of opposite charges in a homogenous magnetic field. Both sets of solutions were obtained my means of a recurrence relation on the coefficients $a_n$ in an anzats  $~\left(\,\sum_na_n \rho^n\,\right)\,\exp(.\,.\,.)\,$ for the wave function.
In~\cite{turbiner94}, the algebraic structure of the Hamiltonian of a system of two electrons in an external oscillator potential was revealed.

In this paper, we consider planar motion of two interacting charged spinless nonrelativistic particles subject to a constant uniform magnetic field. While Turbiner and Escobar-Ruiz~\cite{twocharges,aop1,aop2} addressed the case of a Coulomb interaction between the particles, we here explore other potentials that can permit for the afore-described reduction of the Hamiltonian.
Of a special interest are the potentials
resembling effective interaction between charges in realistic physical settings~\cite{aop2}.

In Section \ref{Sec2}, we briefly discuss the underlying idea of the method, an idea that ensues from the basic properties of representations of Lie groups. In Section \ref{Sec3}, we provide the statement of the problem and carry out some prelusory calculations. In the subsequent sections we derive and explore three reducible potentials never presented in the literature hitherto.

\section{Basics of QES}\label{Sec2}

Let~$\hat{J}_n$ be the generators of an irreducible finite-dimensional representation of the~$SL(2,C)$ group in some vector space:
\begin{equation}
    [\Jnn, \Jnp] = \Jnp \quad,\qquad
    [\Jnn, \Jnm] = - \Jnm \quad,\qquad
    [\Jnp, \Jnm] = 2 \Jnn \quad, \label{generat}
\end{equation}
where~$n = 2j + 1$ is the dimension of the space, and~$j$ is the spin of the representation. These relations are satisfied by the operators defined as
\begin{equation}\begin{aligned}
    \Jnp &= 2 j \rho - \rho^2 \partial_\rho \quad,\\
    \Jnm &= \partial_\rho \quad,\\
    \Jnn &= - j + \rho \partial_\rho \quad.
    \label{2}
\end{aligned}\end{equation}
For each~$n$, we find the corresponding vector space to be
\begin{equation}
    V = \Span \{ 1, \rho, \ldots \rho^{2j} \} \quad.
    \label{3}
\end{equation}
Indeed, these monomials are eigenvectors of~$\Jnn$, while~$\Jnp$ and~$\Jnm$ are the raising and lowering operators. None of these operators can take us outside the representation space~$V$. As a result of this,~$V$ stays invariant under the action of any polynomial function~$\hat{O}$ of the operators~$\hat{J}_n\,$:
\begin{equation}
    \hat{O} \, V \subseteq V \quad.
    \label{4}
\end{equation}
Solving the associated spectral problem
\begin{eqnarray}
\hat{O} \, f(\rho) = - \nu f(\rho) 
\label{spectr}
\label{5}
\end{eqnarray}
over the vector space \eqref{3}
is then equivalent to diagonalisation of an $\,n \times n\,$ matrix. This becomes obvious if one implements the basis vectors~(\ref{3}) and the differential operators~\eqref{2} with column vectors and matrices, correspondingly.

In what follows, we shall be interested in operators of the type
\begin{equation}
\hat{h} = a_{\alpha \beta} \hat{J}^\alpha_n \hat{J}^\beta_n + b_\alpha \hat{J}^\alpha_n \quad,
\label{6}
\end{equation}
because all exactly and quasi-exactly solvable one-dimensionsl Hamiltonians known hitherto can be reduced to this form. In this way, the spectral problem for the operator~$\,\hat{h}\,$,
\begin{eqnarray}
    \left[a_{\alpha \beta} \hat{J}^\alpha_n \hat{J}^\beta_n + b_\alpha \hat{J}^\alpha_n\right]~f(\rho)~=~-~\nu~f(\rho) 
\quad,
\label{h}
\end{eqnarray}
is solvable by purely algebraic means.

\section{Formulation of the problem} \label{Sec3}

We consider a system of two non-relativistic charged spinless particles,~$(e_1, m_1)$ and~$(e_2, m_2)$, moving on a plane orthogonal to a magnetic field~$\bm{B} = B \hat{\bm{z}}$. We limit ourselves to a constant and uniform magnetic field, a restriction which permits us to use the symmetric gauge~$A_{1,2} = \frac{1}{2} \bm{B} \times \bm{\rho}_{1,2}$. The latter circumstance simplifies the problem greatly and allows us to write the Hamiltonian in the form of
\begin{equation}
    \hat{\mathcal{H}} = \frac{(\hat{\bm{p}}_1 - e_1 \bm{A}_1)^2}{2m_1} +
                        \frac{(\hat{\bm{p}}_2 - e_2 \bm{A}_2)^2}{2m_2} +
                        V(\abs{\bm{\rho}_1 - \bm{\rho}_2}) \quad,
\end{equation}
where~$\hbar = c = 1$, while~$\bm{\rho}_{1,2}$ and~$\hat{\bm{p}}_{1,2} = -i \nabla$ are the coordinates and momenta of the particles.

It is natural to switch to the centre of mass reference frame (CMF):
\begin{equation}
    \bm{R} = \mu_1 \bm{\rho}_1 + \mu_2 \bm{\rho}_2,\qquad
    \bm{\rho} = \bm{\rho}_1 - \bm{\rho}_2 \quad.
\end{equation}
Canonically conjugated to them are the CMF and relative momenta:
\begin{equation}
    \hat{\bm{P}} = \hat{\bm{p}}_1 + \hat{\bm{p}}_2,\qquad
    \hat{\bm{p}} = \mu_2\hat{\bm{p}}_1 - \mu_1\hat{\bm{p}}_2\quad,
\end{equation}
where
\begin{equation}
    M = m_1 + m_2,\qquad \mu_i = \frac{m_i}{M}\quad.
\end{equation}
Using these formulae, one can prove that
\begin{equation}
    \op{P} = - i \nabla_{\bm{R}} \quad,\qquad
    \op{p} = - i \nabla_{\bm{\rho}} \quad.\qquad
\end{equation}
The above new coordinates and momenta enable us to define the CMF and relative angular momenta:
\begin{equation}
    \op{L} = \bm{R} \times \op{P} \quad,\qquad
    \op{\ell} = \bm{\rho} \times \op{p} \quad.
\end{equation}
By direct calculation, one can check that the total pseudomomentum~$\op{K}$ and the total angular momentum~$\op{L}^T$ are the integrals of the system:
\begin{gather}
    \op{K} = \op{p}_1 + e_1 \bm{A}_1 +\op{p}_2 + e_2 \bm{A}_2 =
    \op{P} + q \bm{A}_{\bm{R}} + e_c \bm{A}_{\bm{\rho}}\quad,\\
    \op{L}^T = \bm{\rho}_1 \times \op{p}_1 + \bm{\rho}_2 \times \op{p}_2 =
    \op{L} + \op{\ell} \quad,\\
    [\op{K}, \hat{\mathcal{H}}] = [\op{L}^T, \hat{\mathcal{H}}] = 0\quad,
\end{gather}
where
\begin{equation}
    q = e_1 + e_2 \quad,\qquad e_c = \mu_2 e_1 - \mu_1 e_2 = m_r \left(\frac{e_1}{m_1} -\frac{e_2}{m_2} \right) \quad,
\end{equation}
$e_c$ being a coupling charge and~$m_r$ being the reduced mass of the system:
\begin{equation}
    m_r = \frac{m_1 m_2}{M} \quad.
\end{equation}

After a unitary transformation
\begin{equation}
    U  = e^{-i e_c \bm{A}_{\bm{\rho}} \cdot \bm{R}}\quad,
\end{equation}
the Hamiltonian becomes
\begin{equation}
    \hat{\mathcal{H}}' = U^{-1} \, \hat{\mathcal{H}} \, U =
    \frac{(\hat{\bm{P}} - q \bm{A}_{\bm{R}} - 2 e_c \bm{A}_{\bm{\rho}})^2}{2 M} +
    \frac{(\hat{\bm{p}} - q_W \bm{A}_{\bm{\rho}})^2}{2 m_r} +
    V(\abs{\bm{\rho}_1 - \bm{\rho}_2}) \quad,
\end{equation}
where
\begin{equation}
    q_W = e_1 \mu_2^2 + e_2 \mu_1^2 \quad.
\end{equation}
The eigenfunctions of~$\hat{\mathcal{H}}$ and~$\hat{\mathcal{H}}'$ are related via
\begin{equation}
    \Psi' = \Psi e^{i e_c \bm{A}_{\bm{\rho}} \cdot \bm{R}} \quad.
\end{equation}

The two cases of interest are
\begin{itemize}
\item~$e_c = 0$ \quad,
\item~$q = 0$ and~$\op{P} = 0$ \quad.
\end{itemize}
Both lead to similar equations; so here we are going to focus mainly on the first case.

For $\,e_c = 0\,$, the eigenfunctions can be factorised:
\begin{gather}
    \Psi (\bm{R} \,, \bm{\rho}) = \chi(\bm{R}) \, \psi(\bm{\rho})\quad,\\
    \hat{\mathcal{H}} \Psi = (E_R + E_\rho) \Psi \quad.
\end{gather}
The equation for the CMF variable~$\chi$ renders Landau levels (e.g.,~\cite{twocharges}). The equation for the relative motion acquires the form of
\begin{equation}
    \left[ - \frac{\nabla_{\bm \rho}^2}{2 m_r} - \frac{1}{2}\omega_c \hat{\ell}_z +
    \frac{m_r \omega_c^2 \rho^2}{8} + V(\abs{\bm \rho}) - E_{\rho} \right] \psi(\bm{\rho}) = 0 \quad,
\end{equation}
where
\begin{equation}
    \omega_c = \frac{q B}{M} \quad.
\end{equation}
The vanishing of the commutator~$[\hat{\mathcal{H}} \, , \hat{\ell}_z] = 0$ permits further factorisation:
\begin{align}
    \psi({\bm \rho}) &= \zeta(\rho) \Phi(\varphi) \quad, \label{fact}
\end{align}
where~$\Phi(\varphi)$ is the eigenfunction of the relative angular momentum operator~$\hat{\ell}_z$:
\begin{align}
    \Phi(\varphi) &= e^{i s \varphi} \quad,\label{ang}\\
    \hat{\ell}_z \Phi &= s \, \Phi \quad, \qquad s = 0, \pm 1, \pm2 \ldots~~ \label{angeig}
\end{align}
The ensuing equation for~$\zeta$ is:
\begin{equation}
    \left[- \frac{1}{2 m_r} \partial_\rho^{\,2} -
    \frac{1}{2 m_r \rho} \partial_\rho +
    \frac{s^2}{\rho^2} -
    \frac{1}{2}s\, \omega_c  +
    \frac{m_r \omega_c^2 \rho^2}{8} + V(\rho) - E_{\rho} \right] \zeta(\rho) = 0 \quad. \label{zeta1}
\end{equation}

\section{Various potentials}

In~\cite{twocharges}, the case of the Coulomb interaction~$V(\rho) = \dfrac{e_1 e_2}{\rho}$ between two charges was considered. Here we explore other potentials which permit exact solutions.

\subsection{Potential I}

We begin with
\begin{eqnarray}
    V(\rho) =  \frac{e_1 e_2}{\rho} + 
    \frac{\vartheta}{\rho^2} +
    k_1 \,\rho + 
    k_2 \,\rho^2 \quad. \label{pot1}
\label{30}    
\end{eqnarray}
Then the substitution
\begin{equation}
    \zeta_s(\rho) = e^{-\tau \rho^2 - \eta \rho}
    \rho^{\,\xi} p_s(\rho) \label{subs2}
\end{equation}
furnishes the following equation for~$p_s$:
\begin{equation}\begin{gathered}
    \Biggl[ -\rho \partial^2_\rho + \left( 4 \tau \rho^2 +2 \eta \rho - 1 - 2\xi \right) \partial_\rho +
    \frac{1}{4} \rho^3 \left(\omega_c^2 m_r^2 + 8 k_2 m_r - 16\tau^2 \right) +
    \rho^2 \left(2 k_1 m_r - 4 \eta \tau \right) + \\ +
    \rho \left( 4 \tau \{1 + \xi\} - m_r \{s \omega_c+ 2 E_\rho\} - \eta^2 \right) +
    \frac{8 \vartheta m_r - 4 \xi^2 + 4 s^2}{4 \rho}
    \Biggr] p_s =
    -\left[ \epsilon + \eta (1 + 2 \xi) \right] p_s 
    \quad,
    \label{28n}
\end{gathered}\end{equation}
where
\begin{equation}
    \epsilon = 2 m_r e_1 e_1 \quad.
\end{equation}
If the coefficients~$\tau$,~$\eta$ and~$\xi$ are chosen to be
\begin{align}
    \tau = \frac{1}{4} \sqrt{m_r^2 \omega_c^2 + 8 k_2 m_r}\quad, \label{tau}\\ 
    \eta = \frac{k_1 m_r}{2 \tau} = \frac{2 k_1}{\sqrt{\omega_c^2 + \frac{8 k_2}{m_r}}}\quad,\label{eta}\\
    \xi = \sqrt{s^2 + 2 \vartheta m_r} \label{44} \quad.
\end{align}
the equation (\ref{28n}) becomes:
\begin{equation}\begin{gathered}
    \Biggl[ -\rho \partial^2_\rho + \left( 4 \tau \rho^2 +2 \eta \rho - 1 - 2\xi \right) \partial_\rho +
    \rho \left( 4 \tau \{1 + \xi\} - m_r \{s \omega_c+ 2 E_\rho\} - \eta^2 \right) +
    \Biggr] p_s = 
    -\left[ \epsilon + \eta (1 + 2 \xi) \right] p_s
    \quad.
    \label{43}
\end{gathered}\end{equation}
Our next goal is to cast the equation (\ref{43}) into the form (\ref{h}) or, to be more exact, into the form 
\begin{eqnarray}
        \left[ -\hat{J}^0_n\hat{J}^-_n + \hat{J}^+_n - \alpha \hat{J}^-_n + \beta \hat{J}^0_n \right]\, p_s = -\nu \,p_s \quad, \label{generat1}
        \label{38}
\end{eqnarray}
which is the special case of (\ref{h}). We remind that $\,\hat{J}^{-}_n,\,\hat{J}^{+}_n,\,\hat{J}^0_n\,$ are generators of an irreducible  $\,n-$dimensional representation of the $SL(2,C)$ group.


To that end, we insert into (\ref{generat1}) the operators in the differential form, and change the variable as~$c \rho \rightarrow  \rho$:
\begin{equation}\begin{gathered}
    \left[ -\rho \partial^2_\rho + \left(
            c^2 \rho^2 + \frac{1}{2} (n - 2\alpha) + \beta c \rho
        \right) \partial_\rho -
         n c^2 \rho\right] p_s =
    -\left(\nu - \frac{\beta n}{2}\right) c  p_s \label{33} \quad.
\end{gathered}\end{equation}
Equating of the coefficients in~\eqref{43} to those in~\eqref{33} gives us:
\begin{gather}
    \alpha = 1 + 2\xi + \frac{n}{2} \label{alpha1} \quad,\\
    \beta = \frac{2 \eta}{c} = \frac{4 k_1 m_r}{\left( m_r^2 \omega_c^2 + 8 k_2 m_r \right) ^{3/4}} \label{beta1} \quad,\\
    c = 2 \sqrt{\tau} = \sqrt[4]{m_r^2 \omega_c^2 + 8 k_2 m_r} \label{c1} \quad.
\end{gather}

The energy of the relative motion for a fixed~$n$ is:
\begin{equation}
    E_\rho = \frac{4 \tau (\xi + 1) - s \omega_c m_r + c^2 n - \eta^2}{2 m_r} =
    \frac{\omega_c}{2}\left[\sqrt{1+\frac{\textstyle 8 k_2}{\textstyle \omega_c^2 m_r}} (n + 1 + \sqrt{s^2 + 2 \vartheta m_r} ) - s
    - \frac{4 k_1^2}{\left(\omega_c^2 + \frac{\textstyle 8 k_2}{\textstyle m_r}\right)\omega_c m_r}\right]\quad.
\end{equation}
We see that the infinite degeneracy of the energy for $\abs{s}>0$, which was the case in~\cite{twocharges}, does not take place for nonzero $\vartheta$ and $k_2$.

By comparing RHS of~\eqref{43} and~\eqref{33}, we find
the condition
\begin{gather}
    \epsilon + \eta(1 + 2\xi) =
    c  \left( \nu - \frac{\beta n}{2} \right) \quad.
\end{gather}
Inserting therein the expressions~\eqref{eta},~\eqref{beta1}, and~\eqref{c1} for~$\eta$,~$\beta$ and~$c$ from, we arrive at
\begin{gather}
 \epsilon + \frac{2 k_1}{\sqrt{\textstyle \omega_c^2 + \frac{\textstyle 8 k_2}{\textstyle m_r}}} (1 + 2\sqrt{s^2 + 2 \vartheta m_r} ) =
    \sqrt[4]{m_r^2 \omega_c^2 + 8 k_2 m_r}  \left( \nu - \frac{n}{2}
    \frac{4 k_1 m_r}{\left( m_r^2 \omega_c^2 + 8 k_2 m_r \right) ^{3/4}}
    \right) \label{omega1} \quad.
\end{gather}
This equation serves as a constraint upon the permissible values of the parameters entering our problem. (These include the magnetic field, along with the constants in the expression (\ref{30}) for the potential.) Permissible is a set of values of these parameters, for which the equation (\ref{38}), or its equivalent (\ref{33}), has a solution. The values of all these parameters, except one, can be chosen arbitrarily. The remaining parameter will then assume a discrete set of values, to satisfy the quantisation condition (\ref{omega1}).  With the potential fixed, this may be interpreted as `quantisation' of the magnetic field.
This interpretation will work also for the next potential considered.


In case~$k_1$,~$k_2$ and~$\vartheta$ are equal to zero,$\,$\footnote{~Be mindful that vanishing of~$k_1$ entails the vanishing of~$\beta$ and~$\eta$.} the above equation reduces to
\begin{equation}
    \nu^2 = \frac{\epsilon^2}{\omega_c m_r} \quad,
\end{equation}
which coincides with the equation~$(25)$ in~\cite{twocharges}.

At this point, two questions emerge. First, what if the expression under the square root in (\ref{44}) is negative? Second, how to describe the fall on the centre due to the presence of the inverse-square term in the potential? The two issues turn out to be closely connected, and can thus be resolved simultaneously.

For~$\,\rho\rightarrow 0\,$, the equation for~$\,\zeta(\rho)\,$ assumes the form of
\begin{equation}
    \left(\partial_\rho^{\,2} +
    \frac{1}{\rho} \partial_\rho -
    \frac{A}{\rho^2} \right) \zeta(\rho) = 0 \quad,
\end{equation}
where
\begin{equation}
    A = s^2 + 2 \vartheta m_r \quad.
\end{equation}
As is explained in greater detail in~\cite{LL}, no fall towards the centre takes place for~$\,A\geqslant 0\,$. This helps us to observe that the condition
\begin{equation}
    s^2 + 2 \vartheta m_r \geqslant 0
\end{equation}
is quite reasonable, given the physical content of the problem. 

\subsection{Potential II}

Consider the potential of the form
\begin{equation}
    V(\rho) =  \frac{\vartheta}{\rho^2} +
               k_2\, \rho^2 +
               k_4\, \rho^4 +
               k_6\, \rho^6 \quad. \label{pot2}
\end{equation}
A substitution helpful in this case is
\begin{equation}
    \zeta_s(\rho) = e^{-\tau \rho^4-\eta\rho^2}\rho^{\, \xi} p_s(\rho)\quad. \label{subs3}
\end{equation}
Insertion thereof into the formula (\ref{zeta1}) entails the following equation for $\,p_s\,$:
\begin{equation}\begin{gathered}
    \Biggl[ -\rho \partial^2_\rho +
    \left( 
        8 \tau \rho^4 + 4 \eta \rho^2 - 1 -2 \xi
    \right) \partial_\rho +
    \frac{1}{4} \rho^7 \left(
        8 k_4 m_r - 64 \eta \tau
    \right) +
    \frac{1}{4} \rho^5 \left(
        8 k_6 m_r - 64 \tau^2
    \right) + \\ +
    \frac{1}{4} \rho ^3 \left(
        8 k_2 m_r+\omega_c ^2 m_r^2 -16 \left(\eta ^2-2 (\xi +2) \tau \right)
    \right) + 
    \rho  \left(
        4 \eta  (\xi + 1) - m_r (2 E + s \omega_c )
    \right) +
    \frac{8 \vartheta m_r - 4 \xi^2 + 4 s^2}{4 \rho}
    \Biggr] p_s =0 \quad,
\end{gathered}
\end{equation}
wherefrom we find:
\begin{gather}
    \tau = \frac{\sqrt{2 k_6 m_r}}{4} \quad,\label{tau2}\\
    \eta = \frac{k_4 m_r}{8 \tau} = \frac{\sqrt{2 m_r} k_4}{4 \sqrt{k_6}} \quad,\label{eta2}\\
    \xi = \sqrt{s^2 + 2 \vartheta m_r} \label{xi2}\quad.
\end{gather}
Then the equation takes form
\begin{equation}\begin{gathered}
    \Biggl[ -\rho \partial^2_\rho +
    \left( 
        8 \tau \rho^4 + 4 \eta \rho^2 - 1 -2 \xi
    \right) \partial_\rho +
    \frac{1}{4} \rho ^3 \left(
        8 k_2 m_r+\omega_c ^2 m_r^2 -16 \left(\eta ^2-2 (\xi +2) \tau \right)
    \right) + \Biggr. \\ + \Biggl.
    \rho  \left(
        4 \eta  (\xi + 1) - m_r (2 E + s \omega_c )
    \right)
    \Biggr] p_s =0 \quad,
    \label{57}
\end{gathered}\end{equation}
We wish to cast the above equation (\ref{57}) into the standard form (\ref{generat1}). To accomplish this, we substitute the operators in the differential form (the variable in~\eqref{generat1} now being called~$\tilde{\rho}$), and change the variable as~$c \tilde{\rho} \rightarrow  \tilde{\rho}$:
\begin{equation}\begin{gathered}
    \left[ -\tilde{\rho} \partial^2_{\tilde{\rho}} +
    \left(
        c^2 \tilde{\rho}^2 +c \beta \tilde{\rho} -\alpha +\frac{n}{2}
    \right) \partial_{\tilde{\rho}} -
    c^2 n \tilde{\rho} \right] p_s = - c\left(\nu - \frac{\beta n}{2}\right) p_s \label{tilde} \quad.
\end{gathered}\end{equation}

After we change the variable one more time,~$\tilde{\rho} \rightarrow \rho^2$, the equation gets simplified to
\begin{equation}\begin{gathered}
    \left[ -\rho \partial^2_\rho +
    \left(
        2 c^2 \rho ^4 + 2 \beta c \rho^2 -2 \alpha + n +1
    \right) \partial_\rho -
    4 c^2 n \rho ^3 + 2 c ( n \beta - 2 \nu) \rho  \right] p_s =0 \quad.\label{perv}
\end{gathered}\end{equation}
It acquires the needed form if we choose the parameters as:
\begin{gather}
    \alpha = 1 + \xi + \frac{n}{2} \quad,\label{alpha2}\\
    \beta = \frac{\eta}{\sqrt{\tau}} = k_4 \left( \frac{m_r}{8 k_6^3} \right)^{1/4} \quad, \label{beta2}\\
    c = 2 \sqrt{\tau} = (2 k_6 m_r)^{1/4} \label{c2} \quad.
\end{gather}
The energy of the relative motion will be:
\begin{gather}
    E_\rho = \frac{c (n \beta - 2 \nu)}{m_r} +\frac{2 \eta}{m_r} (1 + \xi) -\frac{s \omega_c }{2} \label{E}\quad,
    \end{gather}
while the quantisation condition will read as     
    \begin{gather}
\frac{1}{4} \omega_c^2 m_r^2 = 4 \eta^2 - 4 c^2 n - 8 \tau (\xi + 2) - 2 k_2 m_r \quad,
    \label{omega}
\end{gather}
\begin{equation}
    \frac{1}{4} \omega_c^2 m_r^2 = \frac{m_r k_4^2}{2 k_6} - \sqrt{2 k_6 m_r}\left(4 n + 2 \sqrt{s^2 + 2 \vartheta m_r} + 2\right) - 2 k_2 m_r \quad.
\end{equation}

\section{Potential III}

Finally, let us explore the option
\begin{equation}
    V(\rho) = \frac{l_4}{\rho^4} +
              \frac{l_3}{\rho^3} +
              \frac{l_2}{\rho^2} +
              \frac{l_1}{\rho} -
              k_2 \rho^2 \quad. \label{pot3}
\end{equation}
This time the needed substitution and the equation for~$p_s$ are:
\begin{equation}
    \zeta_s(\rho) = e^{-\frac{\tau}{\rho} - \eta\rho}\rho^{\, \xi} p_s(\rho)\quad,\label{zeta3}
\end{equation}
\begin{equation}\begin{gathered}
    \Biggl[ -\rho^2 \partial^2_\rho +
    \left( 
        2 \eta  \rho ^2-(2 \xi +1) \rho -2 \tau
    \right) \partial_\rho +
    \frac{1}{4}\rho ^4 \left(
        \omega_c^2 m_r^2 - 8 k_2 m_r
    \right) -
    \rho^2 \left(
        2 E m_r + \eta^2 + s \omega_c m_r
    \right) + \\ +
    \rho \left(
        2 \eta \xi + 2 l_1 m_r + \eta
    \right) + 
    \frac{2 l_3 m_r - 2 \xi \tau + \tau }{\rho} +
    \frac{2 l_4 m_r - \tau ^2}{\rho ^2}
    \Biggr] p_s =
    \left(
        \xi^2 - 2 \eta \tau - s^2 -2 l_2 m_r
    \right) p_s
    \quad.\label{21}
\end{gathered}\end{equation}

We compare its coefficients with
\begin{equation}
    \left[ -\hat{J}^- \hat{J}^+_n - 2 \alpha \hat{J}^- + \beta \hat{J}^0_n + 2 \gamma \hat{J}^+_n \right] p_s = -\nu p_s \label{generat3}\quad.
\end{equation}
By substituting operators in the differential form, we get
\begin{equation}
\left[
    -\rho^2 \partial_\rho^2 +
    \left(
        2 \gamma \rho^2 + \rho (\beta + n - 2) - 2 \alpha
    \right) \partial_\rho -
    2 \gamma n \rho
\right] p_s = - \left(\nu - \frac{\beta n}{2} + n\right) p_s \quad.
\end{equation}

From where we find:
\begin{gather}
    \alpha = \tau = \sqrt{2 l_4 m_r} \quad,\\
    \gamma = \eta = -\frac{l_1 m_r \tau}{l_3 m_r + \tau (n + 1)} \quad,\\
    \xi = \frac{1}{2} + \frac{l_3 m_r}{\tau} \quad,\\
    \beta = - \frac{2 l_3 m_r}{\tau} - n \quad.\label{beta3}
\end{gather}

The energy of the relative motion is
\begin{equation}
    E_\rho = - \frac{1}{2} \left(s \omega_c + \frac{\eta^2}{2 m_r}\right) =
    - \frac{1}{2} \left[ s \omega_c + \left(\frac{l_1 \sqrt{2 l_4 m_r}}{2(l_3 m_r + \sqrt{2 l_4 m_r} \{n + 1\})}\right)^2 \right] \quad.
\end{equation}

The first constraint on the parameters of the potential has the form of
\begin{equation}
    2 l_2 m_r - 2 \eta \tau -\xi^2 + s^2 = \nu + n \left(1 - \frac{\beta}{2}\right) \quad
    \nonumber
\end{equation}
or, after we substitute the constants:
\begin{equation}
    2 l_2 m_r - \frac{4 l_1 l_4 m_r^2}{l_3 m_r + \sqrt{2 l_4 m_r}( n+1)} -
    \left( \frac{1}{2} + \frac{l_3 m_r}{\sqrt{2 l_4 m_r}} \right)^2 +
    s^2 = \nu + n \left(1 +\frac{l_3 m_r}{\sqrt{2 l_4 m_r}} + \frac{n}{2} \right)
    \quad.
\end{equation}

In this case, the quantisation of parameters of the potential cannot be reduced to the quantisation of the magnetic field. Instead, we have one more constraint:
\begin{equation}
    \omega_c^2 m_r = 8 k_2 \quad.
\end{equation}

\section{The neutral system}

Let us now discuss the case of~$\,e_1 = - e_2 \equiv e\,$ and show that under certain circumstances it also renders exact solutions. Although in the general case the variables in the unitary-transformed Hamiltonian
\begin{equation}
    \hat{\mathcal{H}}' =
    \frac{(\hat{\bm{P}} - e \bm{B} \times \bm{\rho} )^2}{2 M} +
    \frac{(\hat{\bm{p}} - e (\mu_2 - \mu_1) \bm{A}_{\bm{\rho}})^2}{2 m_r} +
    V(\abs{\bm{\rho}_1 - \bm{\rho}_2}) \quad
\end{equation}
cannot be separated,\footnote{~It is worth noting that in some situations, where the variables in the Hamiltonian cannot be separated, the wave function can nevertheless be factorised.} we nevertheless can consider~$\bm{R}$-independent solutions corresponding to a composite particle at rest:
\begin{equation}
    \Psi_0' (\bm{R}, \bm{\rho}) = \psi(\bm{\rho}) \quad.
\end{equation}
The equation for the relative motion takes the form
\begin{equation}
    \left[ - \frac{\nabla_{\bm \rho}^2}{2 m_r} - \frac{1}{2}\omega_q \hat{\ell}_z +
    \frac{m_r \Omega_q^2 \rho^2}{2} + V(\abs{\bm \rho}) - E_{\rho} \right] \psi(\bm{\rho}) = 0 \quad,
\end{equation}
where
\begin{equation}
    \Omega_q = \frac{e B}{2 m_r}
\end{equation}
and
\begin{equation}
    \omega_q = \frac{e B \abs{\mu_2 - \mu_1}}{m_r} \quad.
\end{equation}
Further steps are analagous to the case of $\,e_c = 0\,$. 
The factorisation~\eqref{fact} leads to the following equation for~$\,\zeta(\rho)\,$:
\begin{equation}
    \left[- \frac{1}{2 m_r} \partial_\rho^{\,2} -
    \frac{1}{2 m_r \rho} \partial_\rho +
    \frac{s^2}{\rho^2} -
    \frac{1}{2}s\, \omega_q  +
    \frac{m_r \Omega_q^2 \rho^2}{2} + V(\rho) - E_{\rho} \right] \zeta(\rho) = 0 \quad,
\end{equation}
which is almost identical to~\eqref{zeta1}. Below we present an inventory of changes in the solutions for the three afore-considered potentials. In what follows,~$\omega_q$ always replaces~$\omega_c$.

\subsection{Potential I}
For the potential of the form~\eqref{pot1}, we again use the substitute~\eqref{subs2}, changing this time~$\eqref{tau}$ to
\begin{equation}
    \tau = \frac{1}{4} \sqrt{4 m_r^2 \Omega_q^2 + 8 k_2 m_r} \quad.
\end{equation}
The equations~\eqref{tau}~--~\eqref{c1} remain unchanged.
The energy of the relative motion and the quantisation condition will look as
\begin{gather}
    E_\rho = \frac{4 \tau (\xi + 1) - s \Omega_q m_r + c^2 n - \eta^2}{2 m_r} =
    \frac{1}{2}\left[\sqrt{4 \Omega_q^2+\frac{8 k_2}{m_r}} (n + 1 + \sqrt{s^2 + 2 \vartheta m_r} ) - s \omega_q
    - \frac{4 k_1^2 \omega_q}{\left(4 \Omega_q^2 + \frac{8 k_2}{m_r}\right) m_r}\right]\quad,\\
    \epsilon + \frac{2 k_1}{\sqrt{4 \Omega_q^2 + \frac{8 k_2}{m_r}}}(1 + 2\sqrt{s^2 + 2 \vartheta m_r}) =
    \sqrt[4]{4 m_r^2 \Omega_q^2 + 8 k_2 m_r}  \left( \nu - \frac{n}{2}
    \frac{4 k_1 m_r}{( 4 m_r^2 \Omega_q^2 + 8 k_2 m_r) ^{3/4}} \right) \quad.
\end{gather}

\subsection{Potential II}
In this case, we still employ the equations~\eqref{subs3}~--~\eqref{c2} and arrive at the same expresssion for the energy:
\begin{gather}
    E_\rho = \frac{c (n \beta - 2 \nu)}{m_r} +\frac{2 \eta}{m_r} (1 + \xi) -\frac{s \omega_q }{2} \label{E2}\quad,
\end{gather}
The quantisation condition assumes the form of
\begin{equation}
    \Omega_q^2 m_r^2 = \frac{m_r k_4^2}{2 k_6} - \sqrt{2 k_6 m_r}\left(4 n + 2 \sqrt{s^2 + 2 \vartheta m_r} + 2\right) - 2 k_2 m_r \quad.
\end{equation}

\subsection{Potential III}
In this situation, the equations~\eqref{zeta3}~--~\eqref{beta3} stay unchanged, and so do the expression for the energy and the first constraint:
\begin{equation}
    E_\rho = - \frac{1}{2} \left(s \omega_q + \frac{\eta^2}{2 m_r}\right) =
    - \frac{1}{2} \left[ s \omega_q + \left(\frac{l_1 \tau}{2(l_3 m_r + \tau \{n + 1\})}\right)^2 \right] \quad,
\end{equation}
\begin{equation}
    2 l_2 m_r - \frac{4 l_1 l_4 m_r^2}{l_3 m_r + \sqrt{2 l_4 m_r}( n+1)} -
    \left( \frac{1}{2} + \frac{l_3 m_r}{\sqrt{2 l_4 m_r}} \right)^2 +
    s^2 = \nu + n \left(1 +\frac{l_3 m_r}{\sqrt{2 l_4 m_r}} + \frac{n}{2} \right)
    \quad.
\end{equation}
The second constrain, however, is now subject to change:
\begin{equation}
    \Omega_q^2 m_r = 2 k_2 \quad.
\end{equation}

\section{Conclusions}

In the presented paper, we extended the class of quasi-exactly solvable problems.
We considered a problem of two charged particles placed in an external magnetic field and interacting with each other through a (generally, non-Coulomb) potential. The first of addresseed potentials was a generalizstion of the potential studied in~\cite{twocharges}. All three considered Hamiltonians resemble those explored in~\cite{turb1988}.
\footnote{~ The difference between our Hamiltonians and those explored in \cite{turb1988} is that in our case there emerges a term due to the presence of the magnetic field. This requires more complicated ans\"atze for the solutions, but the resulting equations in terms of the generators of the~$SL(2,C)$ group are the same.}
They share a common feature: if all of the Hamiltonian's parameters, except one, are fixed, then the exact solutions will exist for an infinite discrete set of values of the remaining parameter. For the first two Hamiltonians, this observation may be written down as `quantisation' of the magnetic field, with the potential's parameters fixed. In the third Hamiltonian, however, the sole permissible value of the magnetic field is unambiguously linked to one of the fixed parameters; so this time it is one of the parameters of the potential that becomes subject to `quantisation'. 

In all three cases, the energy depends on the parameters of the potential and magnetic field, as well as on the magnetic quantum number $s$ and the dimension $n$ of the representation of the generators~$\hat{J}_n$.
Diagolalising a finite $\,n\times n\,$ block of the Hamiltonian, one finds $\,n\,$ energy levels and the corresponding polynomial wave funtions.

\section{Acknowledgements}
The author is grateful to Alexander Turbiner for suggesting this problem and for helpful advise. The author also would like to thank Mikhail Shifman for meticulous reading of the manuscript and very judicious comments that were of great help.

\end{document}